\documentclass[aps,prb,reprint,superscriptaddress,amsmath,longbibliography,amssymb]{revtex4-2}
\usepackage{graphicx}
\usepackage{dcolumn}
\usepackage{bm}
\usepackage{hyperref}
\usepackage{epstopdf}
\usepackage[utf8]{inputenc}
\usepackage[ruled, vlined, linesnumbered]{algorithm2e} 
\SetAlgoLongEnd 
\SetAlgoCaptionSeparator{.} 
\SetAlgoNoEnd 
\SetAlgoNoLine 
\SetNlSty{}{}{:} 
\SetNlSkip{0.4em} 
\SetInd{0.5em}{0.5em} 

\begin{document}
\preprint{APS/123-QED}

\title{Noise-Aware Quantum Architecture Search Based on NSGA-II Algorithm
}

\author{Chenlu Li}
\affiliation{School of Microelectronics, Nanjing University of Science and Technology, Nanjing, Jiangsu 210094, China}
\affiliation{School of Digital and Intelligent Industry, Inner Mongolia University of Science and Technology, Baotou, Inner Mongolia 014010, China}

\author{Hui Zeng}
\email{zenghui@njust.edu.cn}
\affiliation{School of Microelectronics, Nanjing University of Science and Technology, Nanjing, Jiangsu 210094, China}
 
\author{Dazhi Ding}
\email{dzding@njust.edu.cn}
\affiliation{School of Microelectronics, Nanjing University of Science and Technology, Nanjing, Jiangsu 210094, China}
\altaffiliation{Authors to whom correspondence should be addressed: zenghui@njust.edu.cn and dzding@njust.edu.cn}


\begin{abstract}
Quantum architecture search (QAS) has emerged to automate the design of high-performance quantum circuits under specific tasks and hardware constraints. We propose a noise-aware quantum architecture search (NA-QAS) framework based on variational quantum circuit design. By incorporating a noise model into the training of parameterized quantum circuits (PQCs) , the proposed framework identifies the noise-robust architectures. We introduce a hybrid Hamiltonian $\varepsilon$ -greedy strategy to optimize evaluation costs and circumvent local optima. Furthermore, an enhanced variable-depth NSGA-II algorithm is employed to navigate the vast search space, enabling an automated trade-off between architectural expressibility and quantum hardware overhead.  The effectiveness of the framework is validated through binary classification and iris multi-classification tasks under a noisy condition. Compared to existing approaches, our framework can search for quantum architectures with superior performance and greater resource efficiency under a noisy condition.
\end{abstract}

\maketitle

\section{Introduction}
Variational quantum algorithms (VQAs) have demonstrated significant potential  in achieving quantum advantage on noisy intermediate-scale quantum (NISQ) devices\cite{ref1}, particularly for applications in quantum chemistry\cite{REF12}, optimization problems\cite{D2,REF11}, and quantum machine learning\cite{40,16}. A central component of VQAs is the construction of parameterized quantum circuits (PQCs)\cite{ZZC}, which is referred to as the \textit{ansatz}\cite{ref2,ref3,ref4}.  However, manually designing an optimal ansatz remains a challenging task that requires extensive domain expertise\cite{REF13}. While hardware-efficient ansatzes can simplify design, they typically lack flexibility\cite{ref5,REF6}. These predefined circuits generally incorporate redundant gates, which increase both circuit depth and susceptibility to noise.

To overcome these limitations, quantum architecture search (QAS) has been introduced to automate the design of high-performance quantum circuits tailored to specific tasks and hardware constraints\cite{REF7,REF8}. As a discrete optimization problem, QAS traditionally uses heuristic search strategies\cite{6}, including reinforcement learning\cite{3} and evolutionary algorithms(EAs)\cite{LiZhang-854,EA1}. While these search strategies facilitate an effective exploration of the search space, their evaluation incurs a formidable computational overhead. Each candidate quantum architecture should be trained and evaluated repeatedly to determine its ground-truth performance. Within the framework of VQAs, the evaluation process is computationally expensive due to three factors: (1) parameter optimization; (2) gradient computation limited by the barren plateau phenomenon\cite{33}; (3) repeated classical-quantum data exchange. Consequently, developing efficient, low-overhead strategies for  QAS is essential for enhancing the scalability and practical utility of VQAs.

To mitigate the computational overhead of architecture evaluation, classical neural predictor \cite{NA1,NA2} is introduced to estimate the performance of candidate quantum architectures. This method employs regression models to map circuit structures to their expected performance, avoiding individual circuit training. However, training a high-accuracy predictor requires substantial labeled quantum architectures as a dataset, which reintroduces significant computational overhead. Self-supervised learning (SSL)\cite{SSL} has recently been explored to reduce this dependency  by pre-training encoders through graph reconstruction. Nevertheless, such an approach primarily captures structural topological features rather than functional attributes directly correlated with VQAs performance. This limitation constrains the generalization capability of predictors and search efficiency.

The inherent noise of NISQ devices induces various errors, such as bit flips and decoherence. The noise degrades or potentially invalidates the performance of VQAs\cite{REF14,REF15,REF16}. Consequently, a robust QAS framework should be noise-aware. It should incorporate the effects of environmental noise during  architectural exploration to identify circuits that maintain high fidelity in real-world noisy environments. 

Although some studies have addressed noise adaptation, the most important challenge remains: the effective integration of noise models into multi-objective search algorithms that support variable-depth architectures. Such an integration must be achieved without compromising the efficiency of architectural exploration. To overcome these challenges, we propose a noise-aware quantum architecture search (NA-QAS) framework. The three principal contributions of this work are as follows:

Firstly, we introduce a parameter-sharing strategy that couples multiple classical linear layers (supernet) to quantum architectures. For the sampled ansatz, the strategy selects the optimal linear layer for back-propagation. All candidate architectures share a common set of quantum circuit parameters, jointly optimized through a hybrid Hamiltonian parameter-sharing strategy. This strategy substantially reduces the computational overhead of evaluation and effectively prevents premature convergence to local optima.

Secondly, we develop an enhanced NSGA-II evolutionary algorithm for searching variable-depth quantum architectures. The fitness function of algorithm incorporates simulations of multiple practical device noise, such as bit-flip and decoherence channel. The algorithm simultaneously optimizes quantum circuit performance, characterized by the expectation value of the task Hamiltonian, and quantum device costs, including CNOT gate count and quantum circuit depth. This approach enables the automated identification of a Pareto-optimal trade-off between quantum architecture performance, noise robustness, and implementation complexity.

Thirdly, we present an automated framework (NA-QAS) that integrates noise-aware quantum neural networks, the hybrid Hamiltonian parameter-sharing strategy, and multi-objective evolutionary search. The effectiveness of NA-QAS is validated through variational quantum algorithm tasks in simulated noisy environments. The results demonstrate that the proposed framework identifies quantum architectures with superior performance and greater resource efficiency than existing approaches under noisy conditions. 

This article is organized as follows: Section II reviews related work on QAS. Section III details  NA-QAS framework, encompassing search space design, noise-aware quantum neural network model, and enhanced evolutionary search algorithm. Section IV presents experimental results and analysis. Section V concludes the article.

\section{Related Work}
The QAS adapts the methodologies of classical neural architecture search to automate the design of high-performance and resource-efficient PQCs\cite{REF9,REF10}. The QAS framework is characterized by three fundamental components: the search space, the search strategy, and the evaluation strategy\cite{REF}. This section reviews recent advancements across these three dimensions.

\subsection{Search Space}
The search space defines the set of candidate architectures accessible to the search algorithm, typically constrained by hardware-native quantum gate sets and qubit connectivity\cite{D3,D5}. A native random selection of quantum gates often results in "chaotic" circuits resembling random unitary matrices, which are highly susceptible to the barren plateau phenomenon\cite{ML}. To enhance search efficiency and quantum circuit quality, the hierarchical generation method is proposed. Depending on the definition of the search space, different levels of granularity are available for the quantum circuit generation\cite{REF}. Gatewise search spaces incrementally construct circuits by specifying each quantum gate type and position individually. Layerwise search spaces add complete layers composed of multiple quantum gates. Furthermore, blockwise search spaces can be used, where circuits are generated by a set of quantum gates applied to one or multiple qubits. Despite these advancements, most existing search spaces assume a fixed circuit depth, limiting the flexibility required to adapt to varying noise levels in NISQ devices.

\subsection{Search Strategy}

The search strategy encompasses an approach used to explore the search space. As QAS is essentially a discrete optimization problem, heuristic approaches such as  reinforcement learning (RL)\cite{RL4} and evolutionary algorithms (EAs)\cite{EA3} are widely employed. RL-based QAS approaches model the problem as a Markov Decision Process. Agents improve selection strategies through performance-based reward. EA-based QAS approaches encode circuits as individuals, iteratively evolving populations of candidate architectures through genetic operations such as selection, crossover, and mutation\cite{EA4,EA5,EA6}. Although capable of exploring the search space, both approaches demand performance evaluations for numerous candidate architectures, resulting in significant computational overhead. To mitigate computational overhead, differentiable QAS approaches are emerged, relaxing the discrete search space into a continuous and differentiable domain\cite{D3,D5}. 

Bayesian optimization (BO) is a well-known strategy for optimizing black-box functions that are expensive to evaluate\cite{SSL}. It is successfully applied in neural architecture search and also be explored for QAS. BO approximates the performance of candidate architectures using surrogate models, such as Gaussian processes. It directs the search by maximizing an acquisition function. This approach has shown high efficiency in tasks, such as quantum architecture search and entanglement-layer optimization.

Recent research has investigated generative models, including Transformers and generative adversarial networks (GANs), for automated quantum-circuit generation\cite{ML1,ML3}. For instance, generative quantum eigensolver (GQE) employs GPT-like architectures to generate PQCs for quantum simulation, demonstrating the potential of generative models in QAS\cite{REF}. However, effectively integrating noise models into these strategies remains a significant challenge. 

\subsection{Performance Evaluation}
The performance evaluation provides the feedback necessary to guide the search strategy.  The most direct yet costly approach involves fully training each circuit to obtain its true performance. To overcome this, several efficient evaluation strategies have been proposed.

Weight-sharing strategies train a ‘super-circuit’ that contains all candidate sub-circuits. Estimating sub-circuits performance through shared parameters to avoid independent training. Weight-sharing strategies are validated in QAS, though theoretical issues such as performance ranking consistency remain open.

Performance predictors forecast the performance of architectures and thus avoid costly training of each architecture to obtain its performance. Trainable predictors use neural networks to regress performance from quantum architectures, such as convolutional neural networks\cite{17}, recurrent neural networks\cite{RNN} and graph neural networks\cite{GNN}. Training-free predictors rely on proxies such as Clifford noise resilience, expressibility, or path complexity. Both types enable early filtering of poorly performing architectures.

Meta‑learning and transfer learning have been explored to accelerate adaptation across different tasks \cite{M1,M2,M4}. For instance, meta‑learned generators or predictors can transfer knowledge across related tasks, reducing the number of evaluations required for a new task. Nevertheless, existing evaluation strategies often focus on structural features, failing to account for the functional noise-robustness required for practical NISQ applications. 

\section{Methodology}
Solving more complex problems with quantum computing requires greater quantum resources, including an increased number of qubits, greater quantum circuit depth, and additional two-qubit gates. As quantum circuits grow in size, the search space grows exponentially. Exhaustively search strategies are computationally prohibitive. 

To overcome this challenge, we propose a noise-aware quantum architecture search (NA-QAS) framework. Based on variational quantum circuit design, this framework incorporates a noise model during the training of PQCs to foster the development of  noise-robust architectures. We implement a hybrid Hamiltonian parameter-sharing training strategy via supernets, which balances training efficiency with architectural diversity. Finally, a variable-depth NSGA-II algorithm is employed to identify optimal quantum architectures. The framework efficiently explores high-performance quantum architectures under noisy conditions.

\begin{figure*}
        \centering
        \includegraphics[width=1\linewidth]{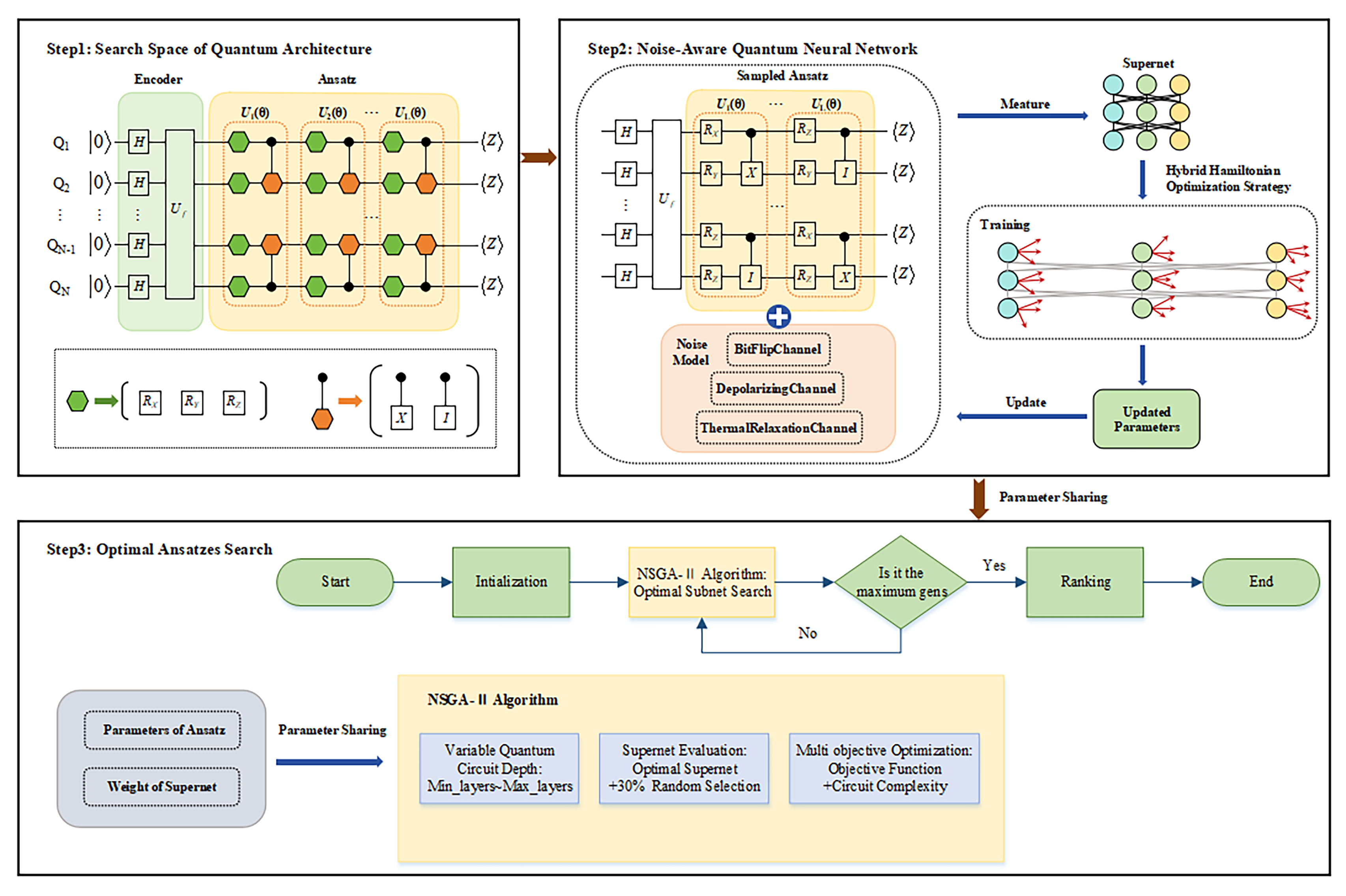}
        \caption{Workflow of NA-QAS framework. In step 1, the unitary $U_f$ refers to the data encoding layer. The search space consists of two components: a rotation-gate combination space and an entanglement-gate combination space. All possible rotation-gates are highlighted by the green hexagon and entanglement-gates are highlighted by the orange hexagon. In step 2, a sampled ansatz is coupled with the noise model, and a hybrid Hamiltonian parameter-sharing strategy is added during the training process. The shared parameters are saved. In step 3, we employ a multi-objective optimization algorithm based on NSGA-II to search for the optimal ansatzes within the search space.}
        \label{FIG.1}
\end{figure*}

As shown in Figure \ref{FIG.1}, the framework consists of three core components: search space of quantum architecture, noise-aware quantum neural network, and multi-objective optimization for QAS.

\subsection{Search Space of Quantum Architecture}
The performance of quantum neural networks depends on the design of quantum architectures. To balance expressibility against computational overhead, we construct a structured and scalable search space for quantum architectures. The search space consists of two components: a rotation-gate combination space and an entanglement-gate combination space.

We define the rotation-gate combination space using the universal single-qubit rotation gates $\{R_x, R_y, R_z\}$ , representing rotations around the $x$, $y$ , and $z$ axes of the Bloch sphere, respectively. In each layer of an \(n\)-qubit ansatz, each qubit can be assigned one of these three gates. The set of all possible single-layer rotation-gate combinations is given by:
\begin{equation}
R_{\text{space}} = \{ R_x, R_y, R_z \}^n
\label{Eq.1}
\end{equation}

For an \(n\)-qubit architecture, each ansatz layer possesses \(3^n\) possible single‑qubit rotation‑gate combinations. We utilize the CNOT gate as the elementary entangling operation. For an \(n\)-qubit ansatz, the set of two‑qubit CNOT connections includes the following pairs:
\begin{equation}
  \text{CNOT}_{\text{space}} = \{ X_{0,1}, X_{0,2}, \ldots, X_{0,n}, \ldots, X_{n-1,n} \}  
  \label{Eq.2}
\end{equation}

where \(X_{ij}\) denotes the CNOT gate with the $ i$-th qubit as the control qubit and the $j$-th qubit as the target qubit, where \(i, j \in [0, n] \text{ and } i \neq j\). To maximize architectural flexibility, each layer may contain any combination of zero to multiple CNOT gates. For an $n$-qubit quantum architecture, each ansatz layer possesses \(2^{n(n-1)/2}\) possible entanglement-gate combinations.

By taking the Cartesian product of the rotation and entanglement subspaces, the complete search space for a single ansatz layer is defined as:  
\begin{equation}
 \label{Eq.3}
S_\text{layer} = R_\text{space}\times \text{CNOT}_\text{space}
\end{equation}

For an $l$-layer quantum architecture, the search space size is\(S_{\text{layer}}^l\). Unlike conventional search strategies that assume fixed depth, the search strategy in this work supports variable-depth architecture. The strategy allows the number of layers $l$ to be dynamically adjusted within the range $[l_{min}, l_{max}]$ during the optimization process. 

\subsection{Noise-Aware Quantum Neural Networks}
We propose a noise-aware quantum neural network model that incorporates multiple noise channels and a hybrid Hamiltonian parameter-sharing strategy. The framework balances training efficiency with architectural robustness by simulating environmental decoherence during the optimization process. 

\subsubsection{Noise Model}
The noise model is implemented by using MindQuantum framework and it encompasses three noise channels that characterize gate operations in real quantum hardware: 
(1) Bit-flip Channel: The bit-flip channel is modeled by applying a Pauli-$X$ operator with probability $p$ following each quantum gate operation. The output quantum state is described by the density matrix:  

\begin{equation}
 \varepsilon_{BF}(\rho) = (1-p)I\rho I + pX\rho X
\label{Eq.4}    
\end{equation}
where \(\rho\) is the density matrix of the input quantum state.

(2) Depolarizing Channel: Analogous to the bit-flip channel, A phase‑flip (Pauli-$Z$ gate) error occurs on a qubit with probability $p$; a combined bit‑phase‑flip (Pauli-$Y$ gate) error also occurs with probability $p$. When $X$, $Y$, and $Z$ gates occur with equal probability, the channel is called a depolarizing channel. This channel is widely used to describe noise from gate operations in real quantum hardware. The output quantum state is given by:
\begin{equation}
    \varepsilon_{DF}(\rho) = (1-p)\rho + \frac{p}{3}(X\rho X + Y\rho Y + Z\rho Z)
    \label{Eq.5}
\end{equation}

(3) Thermal Relaxation Channel: The thermal relaxation channel describes energy relaxation (characterized by time $T_{1}$) and phase relaxation/decoherence (characterized by time $T_{2}$) during a quantum gate operation of duration $T_{g}$. After thermal relaxation, the state is described by:
\begin{equation}
\varepsilon_{TR}(\rho) = \text{tr}_1\left[\Lambda\left(\rho^T \otimes I\right)\right], \quad \Lambda = \begin{pmatrix}
\varepsilon_{T_1} & 0 & 0 & \varepsilon_{T_2} \\
0 & 1-\varepsilon_{T_1} & 0 & 0 \\
0 & 0 & 0 & 0 \\
\varepsilon_{T_2} & 0 & 0 & 1
\end{pmatrix}
\label{Eq.6}
\end{equation}
where \(\varepsilon_{T_1} = e^{-T_g/T_1}\)  and \(\quad \varepsilon_{T_2} = e^{-T_g/T_2}\) denote the probabilities of the qubit remaining unaffected by energy relaxation and decoherence, respectively, within the interval $T_{g}$. \(\quad \Lambda\)  denotes the Choi matrix, representing the complete probability distribution of all possible outcomes of the thermal relaxation channel on the quantum state within the time interval $T_{g}$.

\subsubsection{Hybrid Hamiltonian Parameter-Sharing Strategy}
An ansatz  \(a\) is randomly sampled from the search space and incorporates the noise model to construct a noisy quantum architecture model. To mitigate the high computational cost of training individual candidates, we introduce a parameter-sharing strategy employing multiple supernets. Each supernet is implemented as an independent classical neural network layer:

\begin{equation}
E_k(x) = W_k \cdot x + b_k, \quad k=1,\cdots,K
\label{Eq.7}
\end{equation}
where $K$ denotes the number of supernets, $W_{k}$ and $b_{k}$ represent the weights and biases of the $k$ -th supernet. The initial biases for each supernet are randomly initialized from a normal distribution \(\mathcal{N}(\mu=0, \sigma=0.1)\) , ensuring diversity in initial states.

During each training epoch, the loss function is evaluated in the expectation value of the target operator :
\begin{equation}
    (\theta^*, {k}^*) = \arg\min_{k} \left\langle \psi(\theta,a) \big| \hat{H}_k \big| \psi(\theta,a) \right\rangle
    \label{Eq.8}
\end{equation}
where \(\theta\) are adjustable parameters of quantum gates, \(\hat{H}_k = E_k^\dagger \hat{Z}_k E_k\) represents the target operator for the $k$-th supernet, and \(\hat{Z}\) denotes the measurement operator.

To prevent premature convergence to local optima during training, we introduce the $\varepsilon$-greedy strategy. During each training epoch, the framework randomly selects a supernet for exploration with probability $\varepsilon$ and exploits the current best-performing supernet $k^*$ with probability $1-\varepsilon$. The framework defines a hybrid selection strategy equivalent to optimizing a hybrid Hamiltonian:

\begin{equation}
    \hat{H}_{\text{mix}} = (1-\varepsilon) \hat{H}_{{k}^*} + \varepsilon \hat{H}_{\text{unif}}
    \label{Eq.9}
\end{equation}
where \(\hat{H}_{\text{unif}}\) denotes the uniform hybrid operator.

Similar to conventional VQA approaches, we employ an iterative method to optimize trainable parameters:
\begin{equation}
    \theta^{(t+1)} = \theta^{(t)} - \eta \frac{\partial \left\langle \psi(\theta^{(t)},a) \big| \hat{H}_{\text{mix}} \big| \psi(\theta^{(t)},a) \right\rangle}{\partial \theta^{(t)}}
    \label{Eq.10}
\end{equation}
where \(\eta\) denotes the learning rate. During each training epoch, the supernet selected by the hybrid selection strategy. The weights and biases of supernet updated and saved for subsequent evaluation. The optimized quantum circuit parameters $\theta$ are also stored. These shared parameters avoid retraining each candidate ansatz, substantially reducing the computational overhead.

\subsection{Multi-objective Optimization for Quantum Architecture Search}
After training, we employ an enhanced multi-objective optimization algorithm based on NSGA-II to identify optimal ansatzes within the search space, balancing architecture performance against quantum device cost under noisy conditions.

We formulate the QAS task as a dual-objective optimization problem, simultaneously minimizing the expectation value of the Hamiltonian and the quantum device cost. For a given quantum architecture $\mathcal A$, the fitness function is defined as:
\begin{equation}
    F(\mathcal A) = (\mathcal E(\mathcal A),\mathcal C(\mathcal A))
    \label{Eq.11}
\end{equation}

Here, $\mathcal E(\mathcal A)$ represents the minimum expectation value of the Hamiltonian \(\hat{H}_{\text{task}}\) under optimally tuned parameters $\theta^*$: 
\begin{equation}
    \mathcal E(\mathcal A) = \min_{\theta^*}{\left\langle \psi(\theta^*,\mathcal A) | \hat{H}_{\text{task}} | \psi(\theta^*,\mathcal A)\right\rangle}
    \label{Eq.12}
\end{equation}

$\mathcal C(\mathcal A)$ denotes the quantum device cost, specified as:
\begin{equation}
    \mathcal C(\mathcal A) = \alpha \cdot N_{\text{CNOT}} + \beta \cdot N_{\text{depth}}
    \label{Eq.13}
\end{equation}
where \(N_\text{CNOT}\) denotes the CNOT gate count,  \(N_\text{depth}\)denotes the quantum circuit depth, and $\alpha$ and $\beta$ denote weighting coefficients.

\subsubsection{Improved NSGA-II Algorithm}
The NSGA‑II algorithm is extended to support variable‑depth quantum circuit exploration, as shown in Algorithm\ref{alg:na-qas}. The procedure is as follows:

Step 1: Initialization of the population

An initial population $P_0$ of size $N_{pop}$ is generated. Each individual is represented by a variable‑length gene sequence encoding a quantum architecture:
\begin{itemize}
    \item The gene length \(l \in [l_{\text{min}} , l_{\text{max}}]\) is sampled uniformly from the prescribed interval.
\end{itemize}
\begin{itemize}
    \item Each gene position \(g_i \in \{0, 1, \dots, M-1\}\), where \(M = |\mathcal{S}_{\text{layer}}|\) is the size of the single‑layer search space.
\end{itemize}

Step 2: Fast Fitness Evaluation

For each individual (architecture \(\mathcal{A}_i\)) in the population:

(1) Parameter fine‑tuning: Starting from the shared parameters \(\boldsymbol{\theta}_0\) obtained in pretraining, the architecture is fine‑tuned for \(T\) gradient‑descent steps, yielding an approximately optimal parameter set \(\boldsymbol{\theta}^*_i\).

(2) Performance evaluation: The expectation value \(\mathcal{E}(\mathcal{A}_i)\) of the task Hamiltonian is computed at \(\boldsymbol{\theta}^*_i\).

(3) Cost evaluation: The number of CNOT gates and the circuit depth of  \(\mathcal{A}_i\) are counted, and the resource cost \(\mathcal{C}(\mathcal{A}_i)\) is calculated according to Eq. \ref{Eq.13}.

The fitness vector is defined as:
 \begin{equation}
 \label{Eq.14}
     F(\mathcal{A}_i) = \bigl(\mathcal{E}(\mathcal{A}_i),\, \mathcal{C}(\mathcal{A}_i)\bigr)
 \end{equation}

Step 3: Non‑ dominated Sorting and Crowding‑ distance Calculation

The population $P_t$ is sorted into non‑dominated fronts \(F_1, F_2, \dots\) using fast non‑dominated sorting. Within each front, crowding distances are computed to maintain population diversity.

Step 4: Selection, Crossover, and Mutation

(1) Tournament selection: $k$ individuals are randomly drawn from the parent population, and the fittest among them is retained.

(2) Simulated binary crossover: Selected individuals undergo crossover with probability $p_c$.

(3) Polynomial mutation: Genes are perturbed with probability $p_m$.

(4) Quantum‑specific mutation: To support variable depth, layer insertion, layer deletion, or layer replacement is applied with prescribed probabilities.

Step 5: Elitism and Population Update

The parent population $P_t$ and the offspring population $Q_t$ are combined into $R_t$. Non‑dominated sorting and crowding‑distance assignment are performed on $R_t$. The first $N_{pop}$ individuals are selected to form the new population $P_{t+1}$. The Pareto front $PF_t$ of the current generation is recorded.

 Step 6: Termination and Result Extraction

Steps 2 to 5 are repeated until the maximum generation $G_{max}$ is reached. The algorithm outputs the non‑dominated solutions on the final Pareto front $PF_{G_{max}}$ , or the top $N$ architectures  ranked according to a predefined criterion.

\begin{algorithm}[p]
\SetAlgoLined
\DontPrintSemicolon
\SetKwInOut{Input}{Input}
\SetKwInOut{Output}{Output}
\SetKwProg{Fn}{Function}{}{}

\Input{Population size $N_{pop}$, maximum generations $G_{max}$, depth range $[l_{min}, l_{max}]$,shared parameters $\theta_{shared}$,  Hamiltonian $\hat{H}_{task}$, noise model $\mathcal{N}$}
\Output{Pareto front $PF$ (set of non-dominated solutions)}

\BlankLine
    $P_0 \leftarrow \emptyset$ \tcp*{Initialize population}
    \For{$i \leftarrow 1$ \KwTo $N_{pop}$}{
        $l \leftarrow \text{RandomInt}(l_{min}, l_{max})$ \;
        $A_i \leftarrow \text{GenerateRandomArchitecture}(l)$ \;
        $P_0 \leftarrow P_0 \cup \{A_i\}$ \;
    }
    
    $PF \leftarrow \emptyset$ \tcp*{Global Pareto front}
    
    \For{$t \leftarrow 0$ \KwTo $G_{max}-1$}{
        \tcp{Fast fitness evaluation for each architecture}
        \ForEach{architecture $A \in P_t$}{
            $\theta^* \leftarrow \text{FineTune}(A, \theta_{shared}, T)$ \tcp*{T-step fine-tuning}
                $k^* \leftarrow \arg\min \langle \psi(\theta^*, A) | \hat{H}_\text{mix} | \psi(\theta^*, A) \rangle_{\mathcal{N}}$ \tcp*{Expert selection}
            $\mathcal{E} \leftarrow \langle \psi(\theta^*, A) | \hat{H}_{k^*} | \psi(\theta^*, A) \rangle_{\mathcal{N}}$ \tcp*{Noisy performance}
            $\mathcal{C} \leftarrow \alpha \cdot N_{CNOT}(A) + \beta \cdot N_{depth}(A)$ \tcp*{Resource cost}
            $F(A) \leftarrow (\mathcal{E}, \mathcal{C})$ \tcp*{Two-objective fitness}
        }
        
        $(F_1, F_2, ...) \leftarrow \text{FastNonDominatedSort}(P_t)$ \tcp*{Non-dominated sorting}
        \ForEach{front $F_i$}{
            $\text{CrowdingDistanceAssignment}(F_i)$ \tcp*{Crowding distance}
        }
        
        $Q_t \leftarrow \text{TournamentSelection}(P_t)$ \tcp*{Selection}
        $Q_t \leftarrow \text{SBX}(Q_t, p_c)$ \tcp*{Crossover}
        $Q_t \leftarrow \text{PolynomialMutation}(Q_t, p_m)$ \tcp*{Mutation}
        
        \ForEach{individual $A \in Q_t$}{
            \uIf{with probability $p_{add}$}{
                $\text{InsertLayer}(A)$ \tcp*{Layer insertion}
            }
            \uElseIf{with probability $p_{del}$ \textbf{and} $Depth(A) > L_{min}$}{
                $\text{DeleteLayer}(A)$ \tcp*{Layer deletion}
            }
            \uElseIf{with probability $p_{rep}$}{
                $\text{ReplaceLayer}(A)$ \tcp*{Layer replacement}
            }
        }
        
        $R_t \leftarrow P_t \cup Q_t$ \tcp*{Combine populations}
        $(F_1, F_2, ...) \leftarrow \text{FastNonDominatedSort}(R_t)$ \;
        $P_{t+1} \leftarrow \text{SelectBest}(R_t, N_{pop})$ \tcp*{Elitist selection}
        
        $PF \leftarrow PF \cup F_1$ \tcp*{Update global Pareto front}
        $PF \leftarrow \text{RemoveDominated}(PF)$ \tcp*{Keep only non-dominated}
    }
    \Return{$PF$} \tcp*{Final set of optimal architectures}

\caption{Noise-Aware Quantum Architecture Search (NA-QAS)}
\label{alg:na-qas}
\end{algorithm}

Compared to traditional NSGA-II, this work employs a variable-depth encoding strategy that supports dynamic adjustment of architecture layers. This work overcomes the limitations of fixed-depth architectures, enabling the algorithm to autonomously explore trade-offs between circuit depth and performance. By leveraging the parameter-sharing strategy, it avoids the high computational overhead of retraining each candidate architecture.

\section{Evaluation}
The performance of NA‑QAS is evaluated on two tasks: binary classification and iris multi- classification. Experiments compare the results by different QAS frameworks, under both noisy and noiseless scenarios.

\subsection{Binary Classification}
We first apply NA-QAS to achieve a binary classification task under both noisy and noiseless scenarios. The dataset $D$ contains $n=300$ samples. For each sample $\{x_i, y_i\}$, the feature dimension of input $x_i$ is 3 and the corresponding label $y_i\in \{0,1\}$ is binary.  
At the data preprocessing stage, we split the dataset $D$ into the training set $D_{tr}$, validation set $D_{va}$, and test set $D_{te}$ with size $n_{tr}=100$ ,  $n_{va}=100$ and $n_{te}=100$. The number of qubits is 3 and supernets is $Q=3$ and $K=5$, respectively. The circuit depth is set as $l_{min}=5 , l_{max}=10$. 
\begin{figure}[htbp]
    \centering
    {\includegraphics[width=0.43\textwidth]{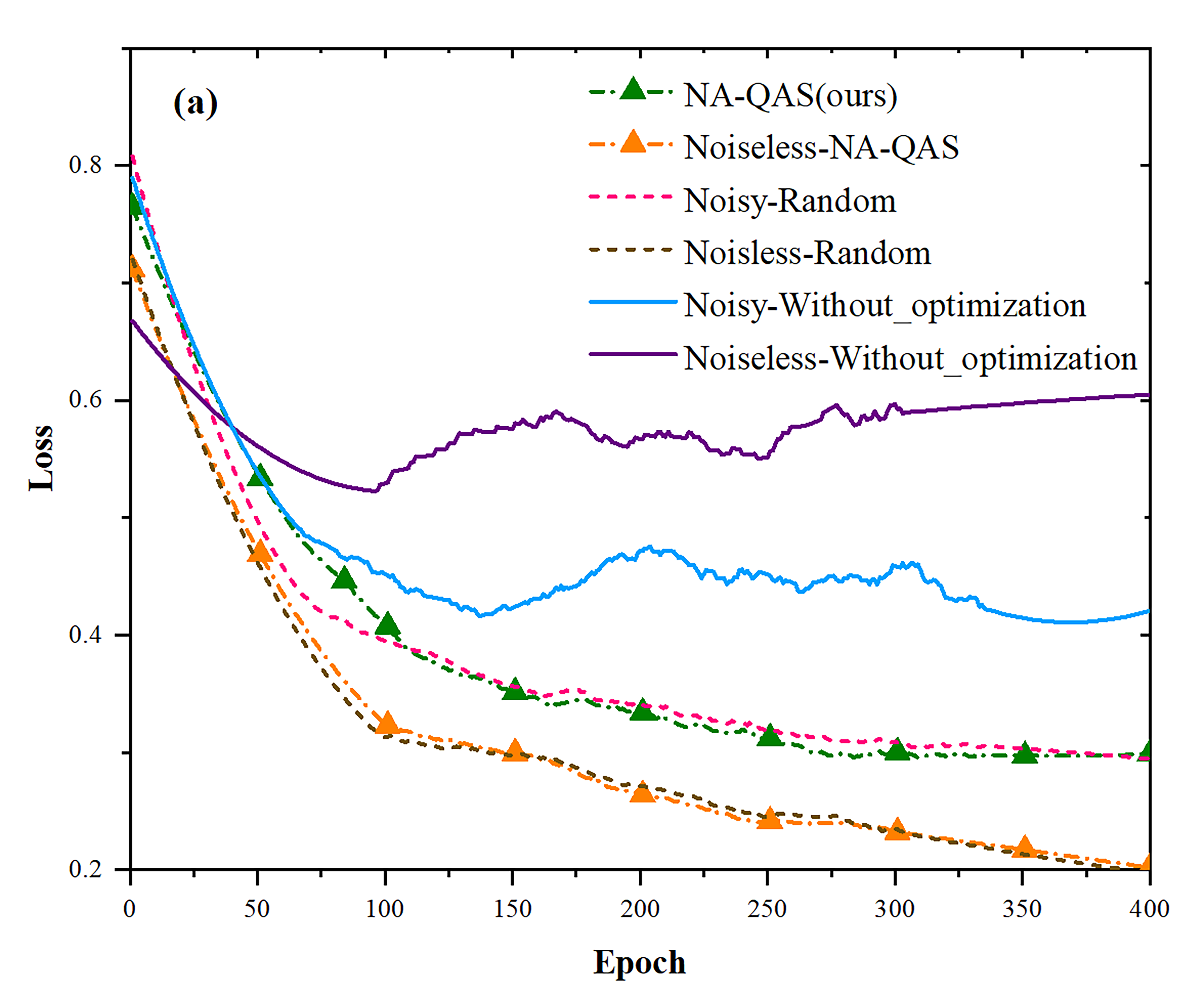}}
    {\includegraphics[width=0.45\textwidth]{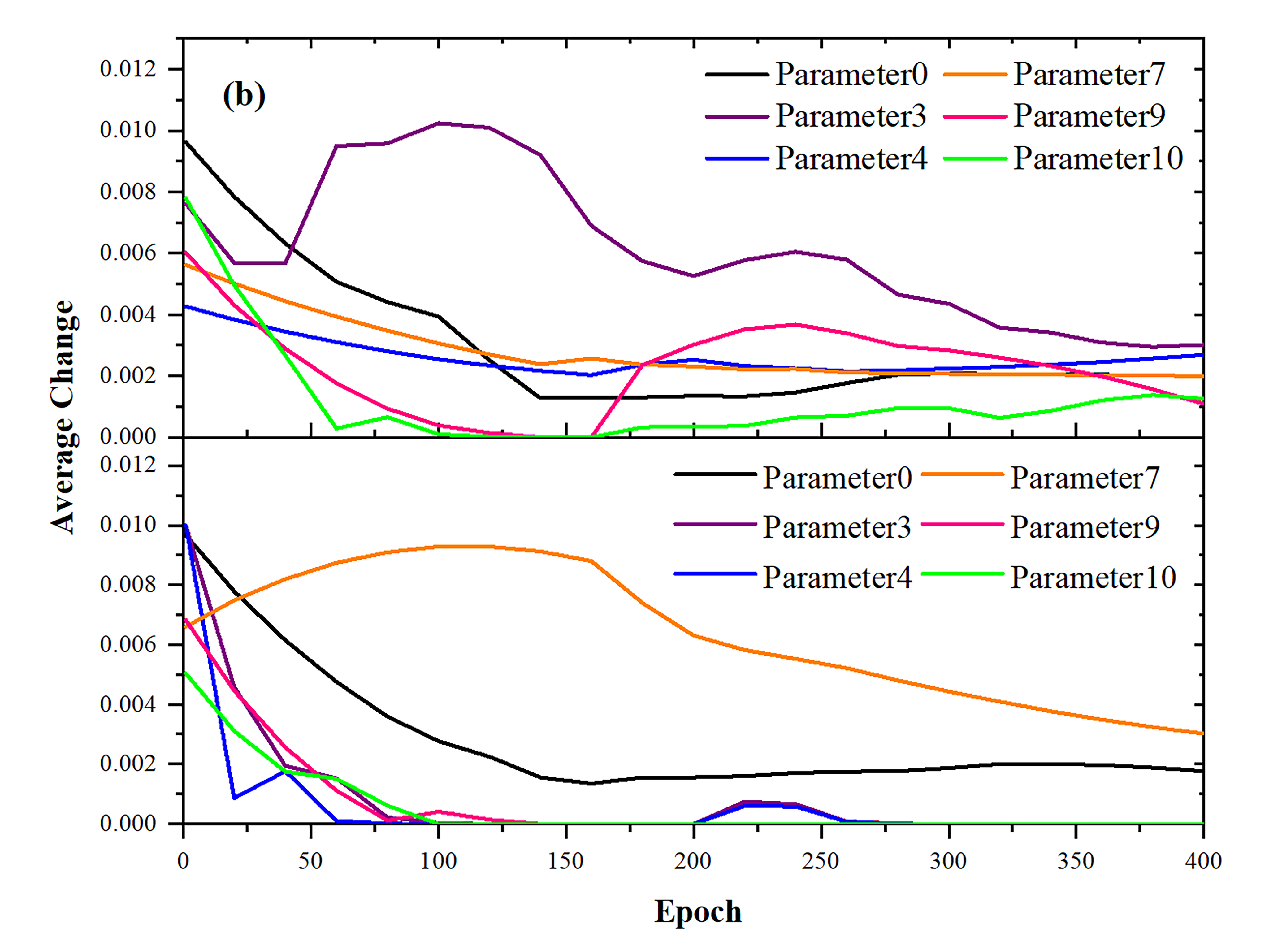}}   
    \caption{Convergence analysis and parameter dynamics for the binary classification task. (a) Evolution of the loss function across training epochs for NA-QAS (ours) compared with random and conventional NSGA-II search methods in both noisy and noiseless scenarios. (b) Average change of trainable ansatz parameters. Top: the parameter evolution with hybrid Hamiltonian parameter-sharing optimization strategy (ours). Bottom: the parameter evolution without hybrid Hamiltonian parameter-sharing optimization strategy.}
    \label{FIG.2}
\end{figure}

Figure \ref{FIG.2}(a) illustrates the convergence behavior of the loss function for different QAS frameworks. The first two curves present comparisons of the proposed NA-QAS under noisy and noiseless scenarios. The third and fourth curves correspond to random search methods that incorporate the proposed optimization strategies (the hybrid Hamiltonian parameter-sharing strategy, noise-aware training, and variable-depth search). The fifth and sixth curves present results from QAS methods based on the conventional NSGA-II algorithm without optimization strategies.

The loss of the two methods (NA-QAS and random search) that use the proposed optimization strategies converge to a stable level as training proceeds, whereas the loss for the method (conventional NSGA-II search) without the strategies fails to converge. This robustness is attributed to the noise-aware training phase, which filters out architectures that are intrinsically sensitive to stochastic errors.

The efficacy of the hybrid Hamiltonian $\varepsilon$-greedy strategy is further validated in Fig. 2(b). The upper panel of Figure \ref{FIG.2}(b) reveals that parameters 9 and 10 exhibit a decline in average change rate to zero around the 150th epoch. However, due to the influence of the proposed hybrid Hamiltonian parameter-sharing optimization strategy, parameters 9 and 10 rapidly escaped the local optima and maintained continuous updates. The lower panel indicates that parameters 3, 4, 9, and 10 all exhibit a decline in average change rate to zero around the 80th epoch, becoming trapped in local optima. Although parameter 3 and parameter 4 briefly update at the 200th epoch, they soon relapse into local optima. For the NA-QAS framework, the average parameter update rate remains non-zero throughout the training epochs, indicating a continuous exploration of the Hilbert space. In contrast, without this strategy, the trainable parameters suffer from premature stagnation, suggesting that the optimizer has collapsed into a local optimum—a common manifestation of the barren plateau phenomenon in noisy landscapes.

\begin{figure}[htbp]
    \centering
      {\includegraphics[width=0.45\textwidth]{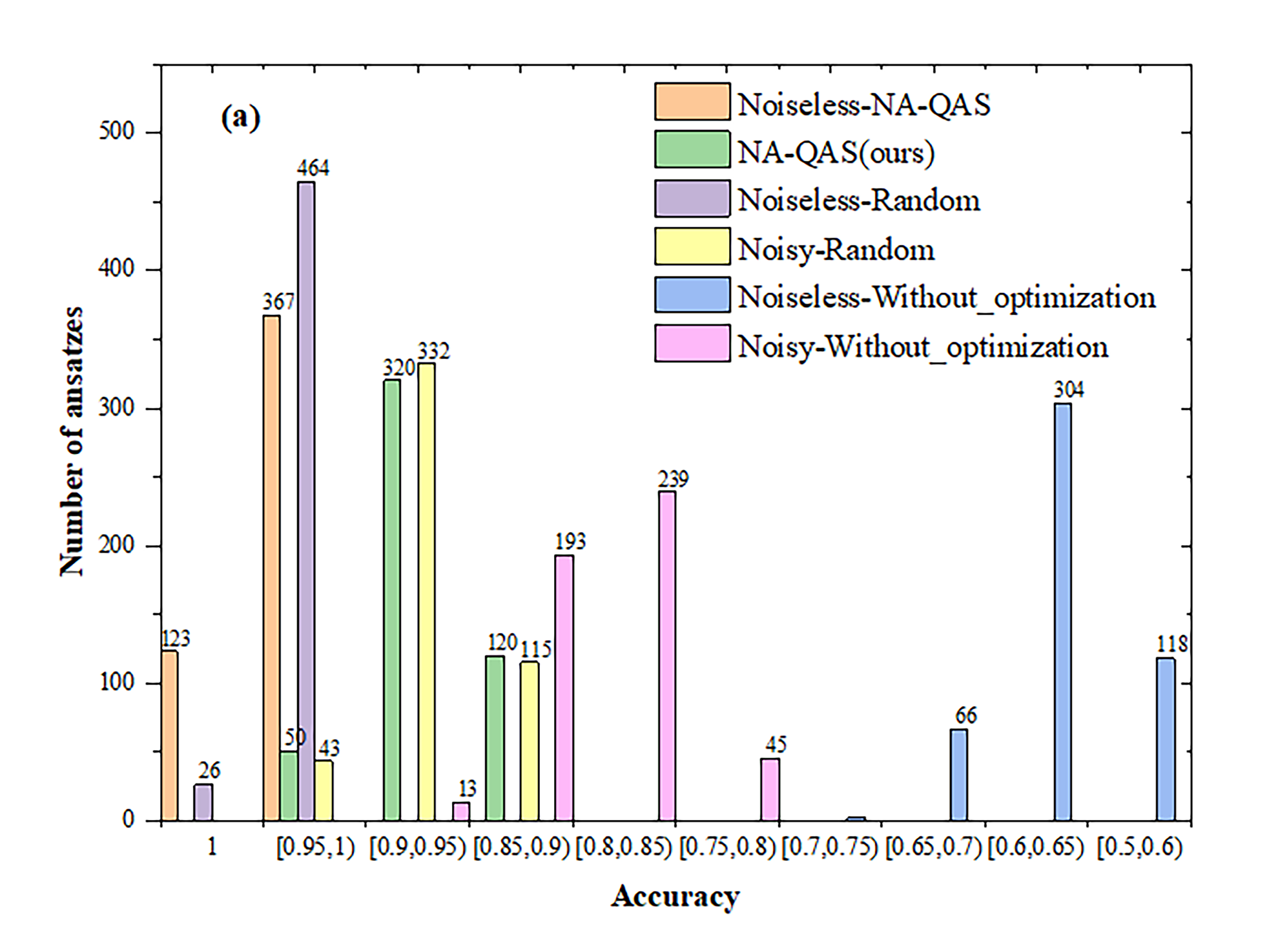}}
    {\includegraphics[width=0.43\textwidth]{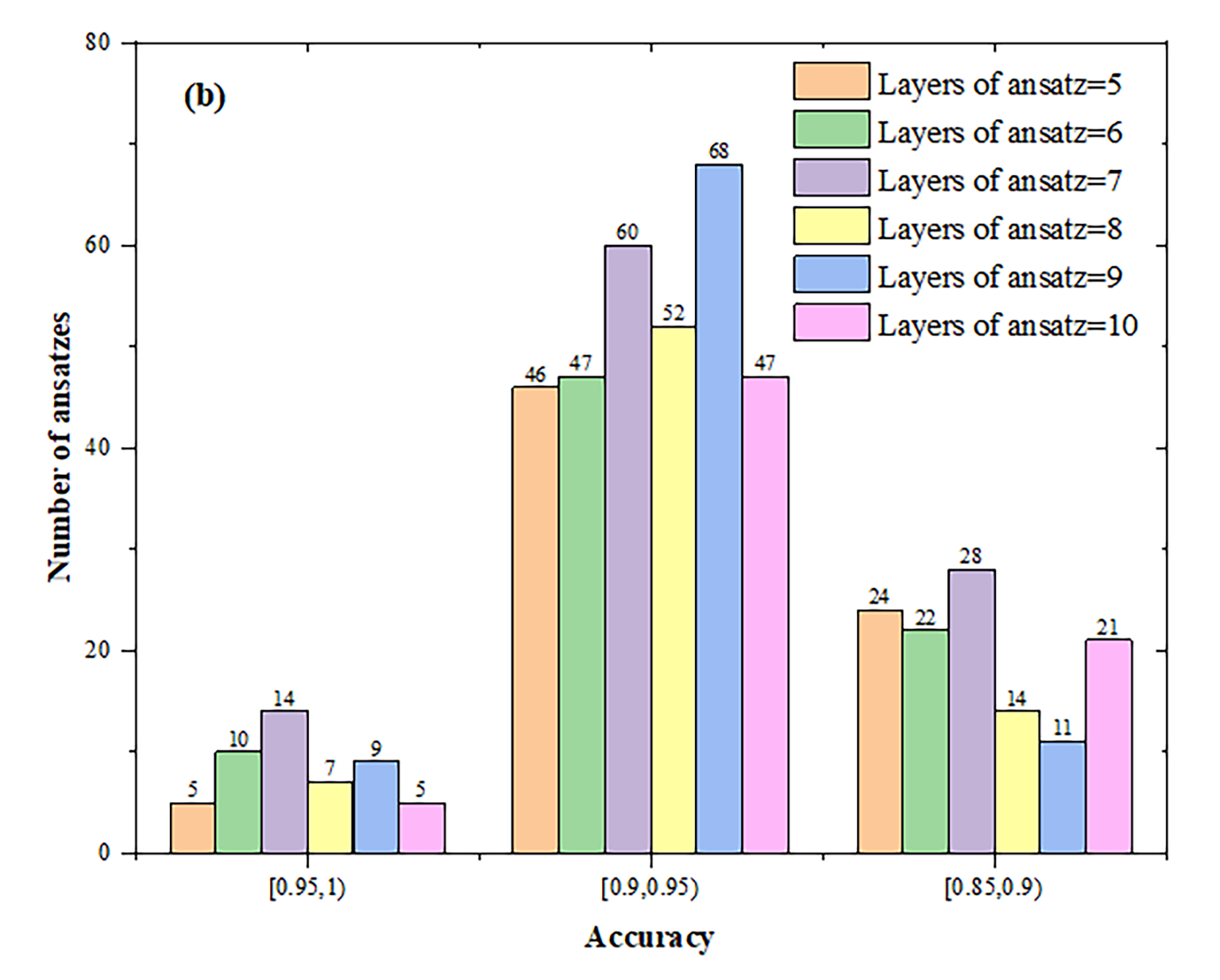}}
    \caption{Statistical distribution of accuracy for the binary classification task. (a) Performance distribution of the ansatzes found by various QAS frameworks, highlighting the superior concentration of high-accuracy solutions generated by NA-QAS. (b) Accuracy distribution across different circuit depths ($l \in [l_{\text{min}}, l_{\text{max}}]$) using NA-QAS, showing the framework’s capability to maintain high expressibility even with reduced quantum resources.}
    \label{FIG.3}
\end{figure}

The statistical distribution of accuracy for the top 490 ansatzes found by different QAS frameworks is depicted in Figure \ref{FIG.3}. As shown in Figure \ref{FIG.3}(a), the methods that use the proposed optimization strategies achieve a clearly higher accuracy than another method. Moreover, the proposed NA‑QAS produces more ansatzes in the high‑accuracy region than other methods. 

Figure \ref{FIG.3}(b) highlights the relationship between circuit depth and performance.  Notably, ansatzes with fewer layers do not show a significant drop in accuracy compared to those with more layers. This result indicates that NA-QAS identifies "sweet-spot" architectures that achieve competitive accuracy with minimal depth, thereby balancing expressibility and quantum circuit depth.

\begin{figure*}[htbp]
    \centering
    \includegraphics[width=1\linewidth]{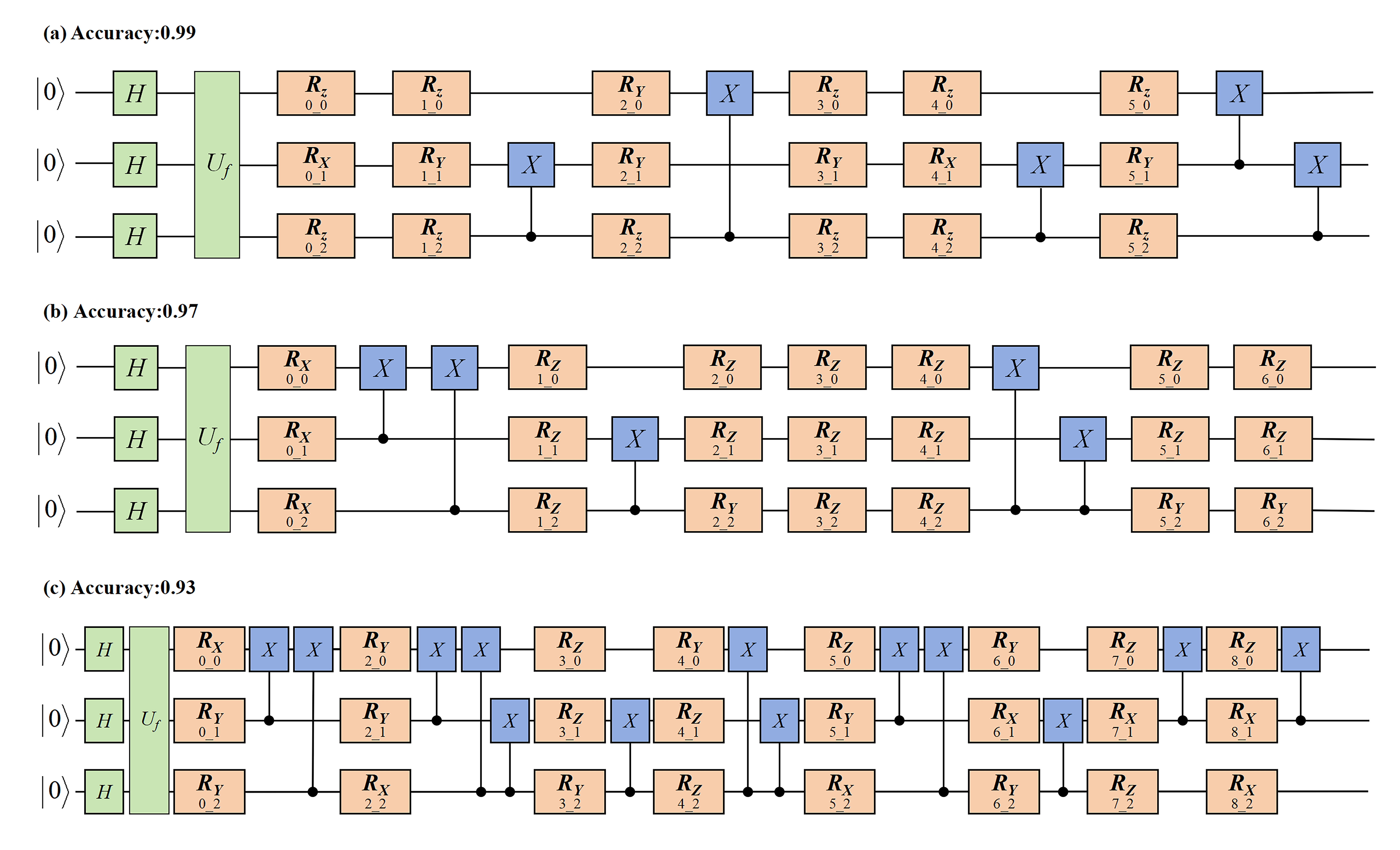}
    \caption{Schematic representation of the optimal quantum architectures identified for the binary classification task under noisy scenarios. (a) The optimal ansatz discovered by the NA-QAS framework, characterized by the minimum CNOT count and circuit depth while achieving peak accuracy. (b) Optimal architecture from random search with optimization strategy. (c) Optimal architecture from evolutionary search without the proposed optimization strategies.}
    \label{FIG.4}
\end{figure*}
Figure \ref{FIG.4} illustrates the optimal quantum architectures obtained through different QAS frameworks under noisy scenarios. Among these, the architecture found by NA-QAS uses the fewest CNOT gates and circuit layers, and has the highest classification accuracy. The NA-QAS framework effectively balances classification accuracy against quantum resource costs.

\subsection{Iris Multi-Classification}
We next apply NA-QAS to achieve an iris multi-classification task under both noisy and noiseless scenarios. The dataset $D$ contains $n=100$ samples. For each sample $\{x_i, y_i\}$, 
the input $x_i$ has four features: sepal length, sepal width, petal length, and petal width.  Each sample is assigned  the corresponding label $y_i$ : $0$ denotes belonging to setosa,  $1$ denotes belonging to versicolor, $2$ denotes belonging to virginica. At the data preprocessing stage, an additional parameter is calculated for each sample using the values of two adjacent features. Consequently, each sample gains three additional parameters, the feature dimension of input $x_i$ is 7. We split the dataset $D$ into the training set $D_{tr}$, validation set $D_{va}$, and test set $D_{te}$ with size $n_{tr}=40$ ,  $n_{va}=30$ and $n_{te}=30$. The number of qubits and supernets is $Q=4$ and $K=5$, respectively. The circuit depth is set as $l_{min}=5 , l_{max}=10$. 

\begin{figure}[htbp]
    \centering
    {\includegraphics[width=0.45\textwidth]{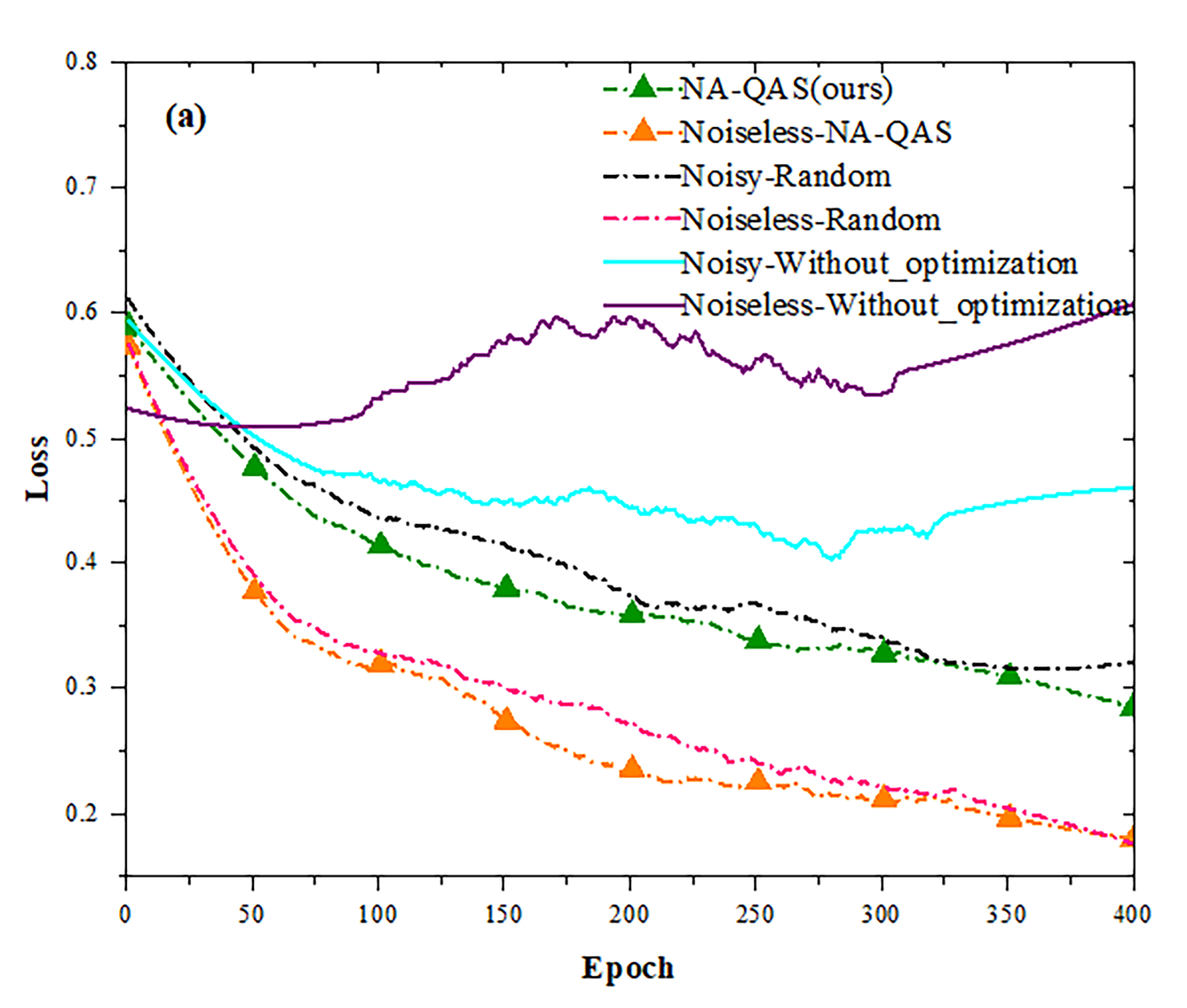}}
    {\includegraphics[width=0.45\textwidth]{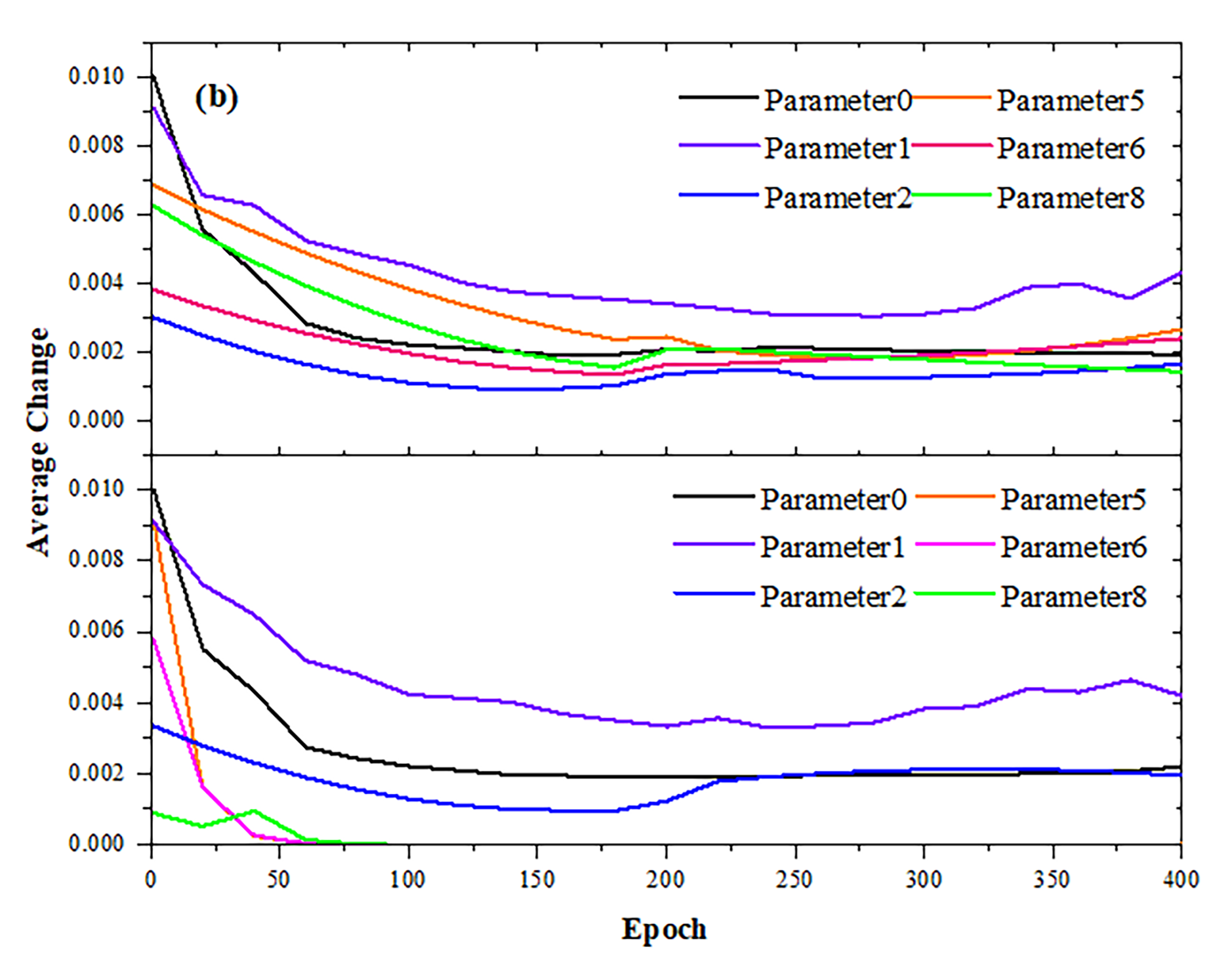}}
    \caption{Convergence analysis and parameter dynamics for the iris multi-classification task. (a) Evolution of the loss function across training epochs for NA-QAS (ours) compared with random and conventional NSGA-II search methods in both noisy and noiseless scenarios. (b) Average change of trainable ansatz parameters. Top: the parameter evolution with hybrid Hamiltonian parameter-sharing optimization strategy (ours). Bottom: the parameter evolution without hybrid Hamiltonian parameter-sharing optimization strategy.}
    \label{FIG.5}
\end{figure}
Figure \ref{FIG.5}(a) illustrates the convergence behavior of the loss function across epochs for different QAS frameworks. The first two curves present comparisons of the proposed NA-QAS under noisy and noiseless scenarios. The third and fourth curves correspond to random search methods that incorporate the proposed optimization strategies (the hybrid Hamiltonian parameter-sharing strategy and variable-depth search). The fifth and sixth curves present results from QAS methods based on the conventional NSGA-II algorithm without optimization strategies.

The loss of the two methods (NA-QAS and random search) that use the proposed optimization strategies converge to a stable level as training proceeds, whereas the loss for the method (conventional NSGA-II search) without the strategies fails to converge.

Figure \ref{FIG.5}(b) illustrates the average change per epoch of six randomly selected trainable parameters in the ansatz under noisy scenario. The upper panel of Figure \ref{FIG.5}(b) indicates that all parameters undergo stable updates throughout the training process. The lower panel indicates that parameter 6 and parameter 8 exhibit a decline in average change rate to approximately zero around the 50th epoch, indicating they have become trapped in local optima. This is also the reason for the non-convergence of the loss in the fifth and sixth curves in Figure \ref{FIG.5}(a).

\begin{figure}
    \centering
    {\includegraphics[width=0.45\textwidth]{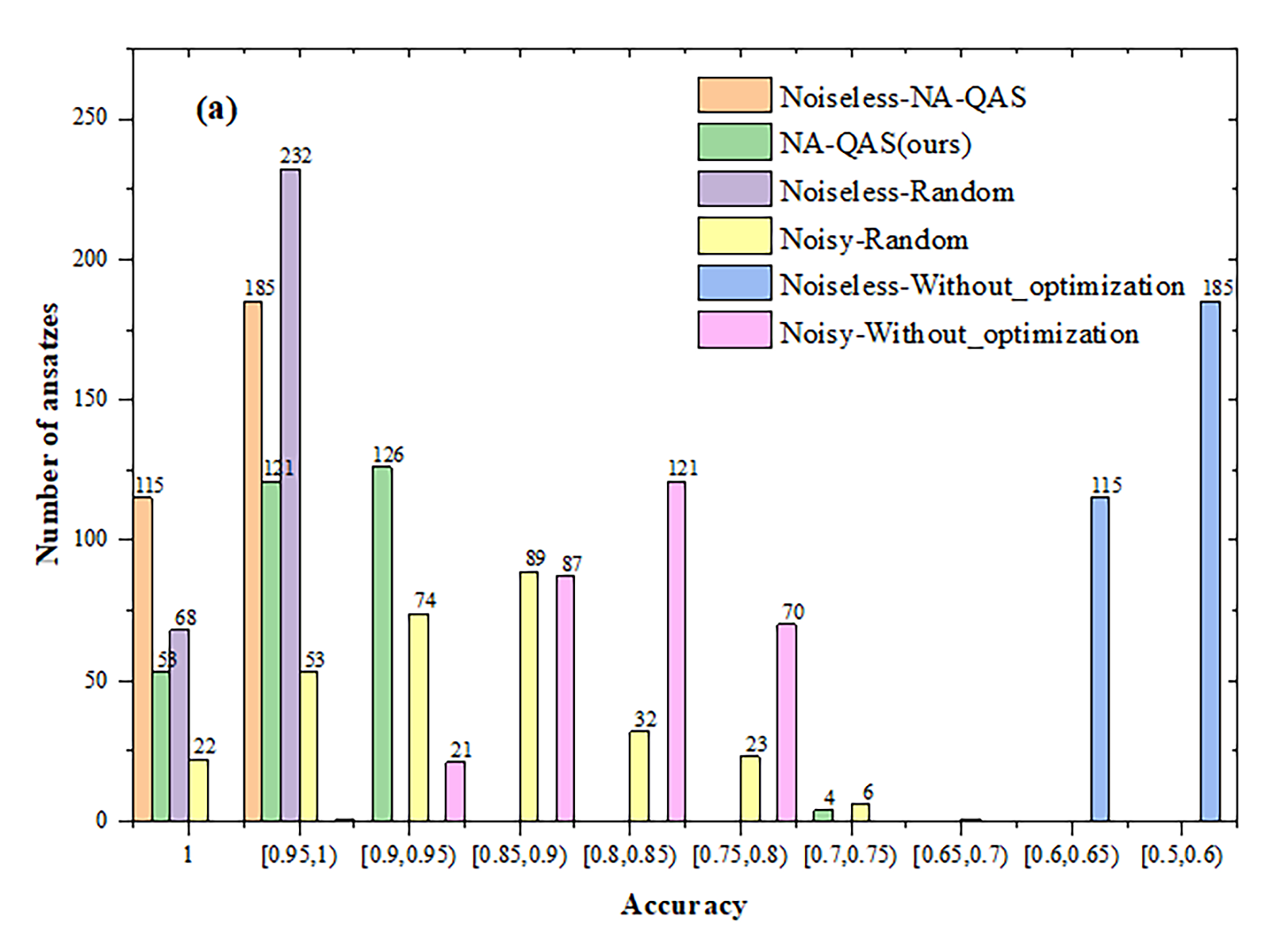}}
     {\includegraphics[width=0.45\textwidth]{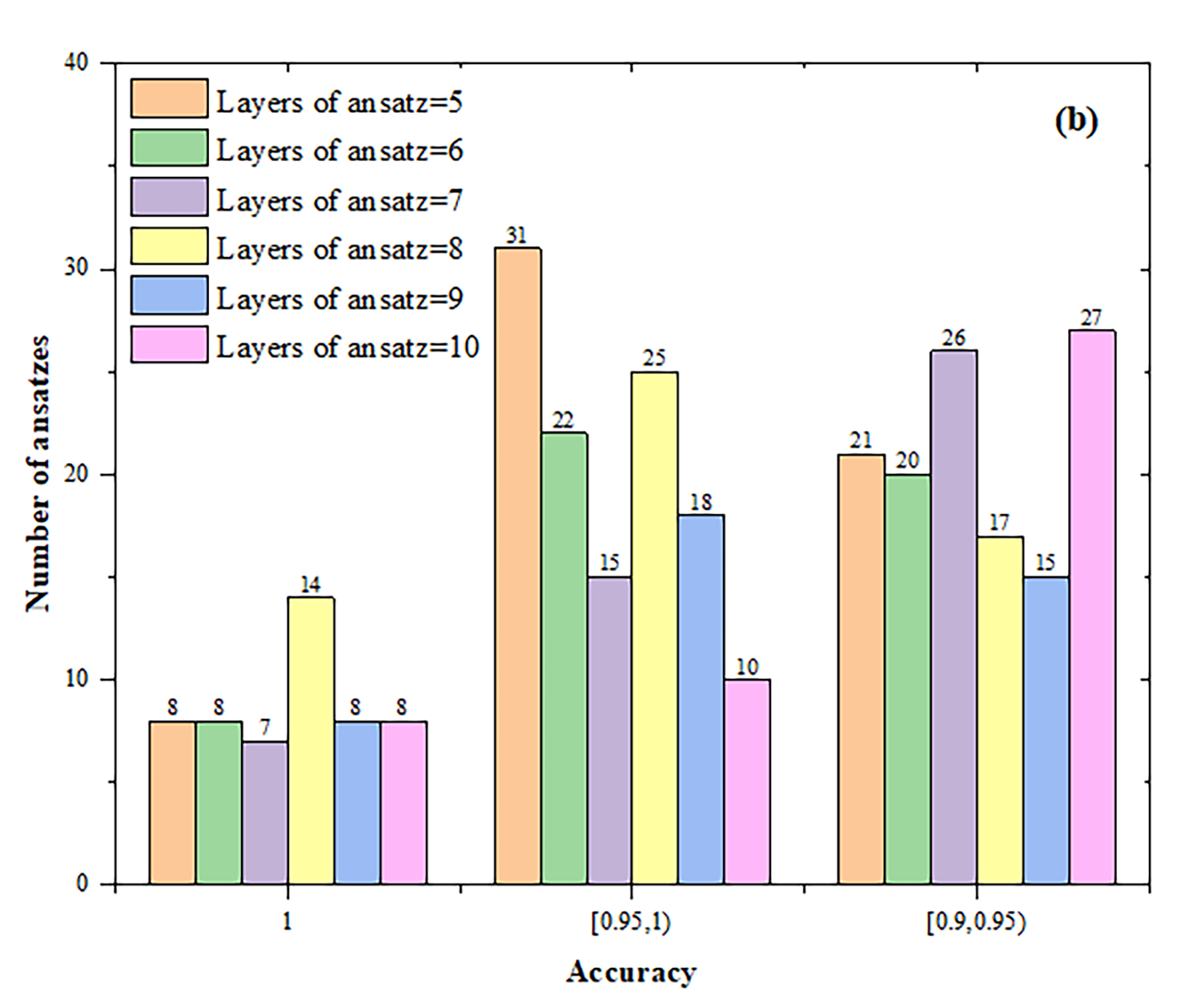}}
    \caption{Statistical distribution of accuracy for the binary classification task. (a) Performance distribution of the ansatzes found by various QAS frameworks, highlighting the superior concentration of high-accuracy solutions generated by NA-QAS. (b) Accuracy distribution across different circuit depths ($l \in [l_{\text{min}}, l_{\text{max}}]$) using NA-QAS, showing the framework’s capability to maintain high expressibility even with reduced quantum resources.}
    \label{FIG.6}
\end{figure}
Figure \ref{FIG.6}(a) presents the statistical distribution for the top 300 ansatzes found by different QAS frameworks for the iris multi-classification task. The methods that use the proposed optimization strategies achieve a clearly higher accuracy than another method. The architectures identified by NA-QAS are concentrated in the high-fidelity region ($>95\%$ accuracy), whereas random search results in a wide, suboptimal variance. 

Figure \ref{FIG.6}(b) compares the accuracy distributions of ansatzes with different numbers of layers, all obtained with NA‑QAS. Ansatzes with fewer layers do not show a significant drop in accuracy compared to those with more layers. This result indicates that NA-QAS successfully balances expressibility and quantum circuit depth.

\begin{table*}
\caption{Performance comparison for optimal quantum architectures identified by different QAS frameworks. The benchmarks were conducted under simulated noisy conditions, including bit-flip, depolarizing, and thermal relaxation channels. The "Accuracy" column denotes the validation accuracy on the Iris classification task, while "CNOT Count" and "Circuit Depth" represent the primary metrics for quantum resource overhead.}
\label{Table-1}
\begin{ruledtabular}
\begin{tabular}{ccccc}
Method& Qubit& Accuracy& CNOT Count& Circuit Depth\\ 
\hline
NA-QAS(ours)& 4& 1& 26& 8
\\
Random-Search(with Optimization Strategies)& 4& 1& 32& 9
\\
Evolutionary-Search
(without Optimization Strategies)& 4& 0.93& 18& 10\\
\end{tabular}
\end{ruledtabular}
\end{table*}
 Table \ref{Table-1} presents the optimal quantum architectures obtained through three different QAS frameworks. At equivalent classification accuracy, the quantum circuit architecture identified by NA-QAS employs fewer CNOT gates and circuit layers. Although Evolutionary-Search (without optimization strategies) employs even fewer CNOT gates, its classification accuracy is considerably lower. NA-QAS framework effectively balances classification accuracy against quantum resource cost.
 
\begin{table*}
\caption{Performance comparison of the NA-QAS framework and existing QAS architectures in noisy conditions. The benchmarks illustrate the superiority of our framework in terms of classification accuracy and resource efficiency under noise channels.}
\label{Table-2}
\begin{ruledtabular}
\begin{tabular}{ccccccc}
 Task&Method&  Stategy&Qubit& Accuracy&  Two-Qubit Gate Type&Two-Qubit Gate Count\\ 
\hline
 Binary Classification&NA-QAS(ours)&  Evolutionary Search&3& 0.99&  CNOT&5\\
 Binary Classification&QuantumNAS&  Evolutionary Search&4& 0.96&  CU&9\\
 Iris Multi-Classification& NA-QAS(ours)&  Evolutionary Search&4& 1&  CNOT&26\\
 Iris Multi-Classification&Enhanced-QAS&  Random Search&4& 0.97&  CZ&4\\
\end{tabular}
\end{ruledtabular}
\end{table*}
As shown in Table\ref{Table-2}, for binary classification tasks, NA-QAS achieves the accuracy of 0.99 using only 3 qubits and 5 CNOT gates. In comparison, QuantumNAS\cite{quantumNAS} requires 4 qubits and 9 CU gates to achieve a lower accuracy of 0.96. This significant reduction in two-qubit gate count (nearly 45\%) demonstrates that the variable-depth evolutionary search effectively prunes redundant operations that would otherwise exacerbate noise-induced errors. In the more challenging Iris classification task, NA-QAS reaches the performance upper bound with a perfect accuracy of 1. Although Enhanced-QAS\cite{E-QAS} utilizes fewer two-qubit gates (4 CZ gates), its accuracy is limited to 0.97.

\section{Conclusion}
In this work, we have proposed a noise-aware quantum architecture search (NA-QAS) framework based on variational quantum circuit design. The framework incorporates a noise model during the training of parameterized ansatzes and leverages a supernet-based parameter-sharing strategy. By employing a hybrid Hamiltonian $\varepsilon$-greedy optimization strategy, the proposed method effectively circumvents the local optimum problem while maintaining a robust balance between training efficiency and architectural diversity. Furthermore, we utilized an enhanced NSGA-II algorithm that supports variable-depth search to identify optimal quantum architectures. Experiments on binary classification and iris multi-classification tasks demonstrate that NA-QAS efficiently discovers high-performance, noise-robust ansatzes. The identified optimal quantum architectures strike a practical balance between expressibility and quantum resource overhead, offering a scalable path for VQA deployment on NISQ-era hardware .

\begin{acknowledgments}
This work was supported by the National Natural Science Foundation of China (No. 62371238), and supported by the Key Project of Science Foundation of Jiangsu Province (No. BK20243046).
\end{acknowledgments}

\bigskip
\textbf{DATA AVAILABILITY}
\par
The data that support the findings of this article are not publicly available upon publication because it is not technically feasible and/or the cost of preparing, depositing, and hosting the data would be prohibitive within the terms of this research project. The data are available from the authors upon reasonable request.

\bigskip
\textbf{ORCID iDs}\\
Hui Zeng https://orcid.org/0000-0002-7657-6714\\
Dazhi Ding https://orcid.org/0000-0001-8522-6233\\

\newpage
\bibliography{reference}

@article{
LiZhang-854,
   Author = {Altares Lopez, Sergio and Ribeiro, Angela and Garcia Ripoll, Juan Jose},
   Title = {Automatic design of quantum feature maps},
   Journal = {Quantum Sci. Technol.},
   Volume = {6},
   pages = {045015},
   DOI = {10.1088/2058-9565/ac1ab1},
   Year = {2021} }

@article{
ML,
   Author = {Zhang, Fei and Li, Jie and He, Zhimin and Situ, Haozhen},
   Title = {Learning the Expressibility of Quantum Circuit Ansatz Using Transformer},
   Journal = {Adv. Quantum Technol.},
   Volume = {8},
   issue = {6},
   pages = {1002},
   DOI = {10.1002/qute.202400366},
   Year = {2025} }

@article{
ML3,
   Author = {Suzuki, Yudai and Sakuma, Rei and Kawaguchi, Hideaki},
   Title = {Light-Cone Feature Selection for Quantum Machine Learning},
   Journal = {Adv. Quantum Technol.},
   Volume = {8},
   issue = {6},
   pages = {1002},
   DOI = {10.1002/qute.202400647},
   Year = {2025} }

@article{ 
ref1,
   Author = {Situ, Haozhen and Li, Zhengjiang and He, Zhimin and Li, Qin and Shi, Jinjing},
   Title = {AutoML-driven optimization of variational quantum circuit},
   Journal = {Inf. Sci.},
   Volume = {717},
   pages = {122272},
   DOI = {10.1016/j.ins.2025.122272},
   Year = {2025} }

@inproceedings{
ref2,
   Author = {Chen, Samuel Yen Chi},
   Title = {Evolutionary Optimization for Designing Variational Quantum Circuits with High Model Capacity},
   BookTitle = {2025 IEEE Symposium for Multidisciplinary Computational Intelligence Incubators, MCII Companion},
   Address= {Trondheim, NORWAY},
   Year = {2025} }

@article{
ref3,
   Author = {Yao, Yuhan and Hasegawa, Yoshihiko},
   Title = {Avoiding barren plateaus with entanglement},
   Journal = {Phys. Rev. A},
   Volume = {111},
   issue = {2},
   pages = {022426},
   DOI = {10.1103/PhysRevA.111.022426},
   Year = {2025} }

@inproceedings{
ref4,
   Author = {Xiang, Debin and Jiang, Qifan and Lu, Liqiang and Tan, Siwei and Yin, Jianwei},
   Title = {Choco-Q: Commute Hamiltonian-based QAOA for Constrained Binary Optimization},
   BookTitle = {2025 IEEE International Symposium on High Performance Computer Architecture, HPCA},
   Address= {Las Vegas, NV},
   Pages = {275-289},
   Year = {2025} }

@article{
ref5,
   Author = {Du, Yuxuan and Tu, Zhuozhuo and Yuan, Xiao and Tao, Dacheng},
   Title = {Efficient Measure for the Expressivity of Variational Quantum Algorithms},
   Journal = {Phys. Rev. Lett.},
   Volume = {128},
   issue = {8},
   pages = {080506},
   DOI = {10.1103/PhysRevLett.128.080506},
   Year = {2022} }

@article{
16,
   Author = {Dong, Daoyi and Chen, Chunlin and Chu, Jian and Tarn, Tzyh-Jong},
   Title = {Robust Quantum-Inspired Reinforcement Learning for Robot Navigation},
   Journal = {IEEE-ASME Trans. on Mechatronics},
   Volume = {17},
   Pages = {86-97},
   DOI = {10.1109/TMECH.2010.2090896},
   Year = {2012} }

@article{
REF6,
   Author = {Cerezo, M. and Sone, Akira and Volkoff, Tyler and Cincio, Lukasz and
   Coles, Patrick J.},
   Title = {Cost function dependent barren plateaus in shallow parametrized quantum
   circuits},
   Journal = {Nat. Commun.},
   Volume = {12},
   pages = {1791},
   DOI = {10.1038/s41467-021-21728-w},
   Year = {2021} }

@article{
REF7,
   Author = {Chen, Xinyu and Zhu, Mingqiang and Cheng, Xueyun and Guan, Zhijin and Feng, Shiguang and Zhu, Pengcheng},
   Title = {Nearest-neighbor synthesis of controlled-NOT circuits on general quantum architectures},
   Journal = {Phys. Rev. A},
   Volume = {111},
   pages = {062617},
   DOI = {10.1103/13m1-y3br},
   Year = {2025} }

@article{
REF8,
   Author = {Bi, Xiao Yu and Yu, Yi Ming and Chen, Ye Hong and Zhong, Zhi Rong},
   Title = {General-purpose quantum architecture search based on deep reinforcement
   learning},
   Journal = {Phys. Rev. A},
   Volume = {112},
   pages = {052409},
   DOI = {10.1103/7rc4-p446},
   Year = {2025} }

@article{
17,
   Author = {Rapp, Frederic and Kreplin, David A. and Huber, Marco F. and Roth, Marco},
   Title = {Reinforcement learning-based architecture search for quantum machine learning},
   Journal = {Machine Learning-Science and Technology},
   Volume = {6},
   pages = {0150411},
   DOI = {10.1088/2632-2153/adaf75},
   Year = {2025} }

@article{
REF9,
   Author = {Bharti, Kishor and Cervera Lierta, Alba and Kyaw, Thi Ha and Haug,
   Tobias and Alperin-Lea, Sumner and Anand, Abhinav and Degroote, Matthias
   and Heimonen, Hermanni and Kottmann, Jakob S. and Menke, Tim and Mok,
   Wai-Keong and Sim, Sukin and Kwek, Leong-Chuan and Aspuru-Guzik, Alan},
   Title = {Noisy intermediate-scale quantum algorithms},
   Journal = {Rev. Mod. Phys.},
   Volume = {94},
   pages = {015004},
   DOI = {10.1103/RevModPhys.94.015004},
   Year = {2022} }

@article{
REF10,
   Author = {Su, Junjian and Fan, Jiacheng and Wu, Shengyao and Li, Guanghui and Qin, Sujuan and Gao, Fei},
   Title = {Topology-driven quantum architecture search framework},
   Journal = {Sci. China Inf. Sci.},
   Volume = {68},
   pages = {1805078},
   DOI = {10.1007/s11432-024-4486-x},
   Year = {2025} }

@article{
SSL,
   Author = {He, Zhimin and Chen, Hongxiang and Zhou, Yan and Situ, Haozhen and Li, Yongyao and Li, Lvzhou},
   Title = {Self-supervised representation learning for Bayesian quantum architecture search},
   Journal = {Phys. Rev. A},
   Volume = {111},
   pages = {032403},
   DOI = {10.1103/PhysRevA.111.032403},
   Year = {2025} }

@article{
6,
   Author = {Lipardi, Vincenzo and Dibenedetto, Domenica and Stamoulis, Georgios and Winands, Mark H. M.},
   Title = {Quantum Circuit Design using a Progressive Widening Enhanced Monte Carlo Tree Search},
   Journal = {Adv. Quantum Technol.},
   volume = {8},
   issue = {11}, 
   pages = {1002},
   DOI = {10.1002/qute.202500093},
   Year = {2025} }

@article{
NA1,
   Author = {Li, Yangyang and Liu, Guanlong and Zhao, Peixiang and Shang, Ronghua and Jiao, Licheng},
   Title = {Balanced neural architecture search},
   Journal = {Neurocomputing},
   Volume = {594},
   pages = {127860},
   DOI = {10.1016/j.neucom.2024.127860},
   Year = {2024} }

@article{
M1,
   Author = {He, Zhimin and Chen, Chuangtao and Li, Zhengjiang and Situ, Haozhen and Zhang, Fei and Zheng, Shenggen and Li, Lvzhou},
   Title = {A meta-trained generator for quantum architecture search},
   Journal = {EPJ Quantum Technology},
   Volume = {11},
   pages = {441},
   DOI = {10.1140/epjqt/s40507-024-00255-9},
   Year = {2024} }

@article{
ML1,
   Author = {Kundu, Akash and Sarkar, Aritra and Sadhu, Abhishek},
   Title = {KANQAS: Kolmogorov-Arnold Network for Quantum Architecture Search},
   Journal = {EPJ Quantum Technology},
   Volume = {11},
   pages = {761},
   DOI = {10.1140/epjqt/s40507-024-00289-z},
   Year = {2024} }

@article{
REF11,
   Author = {Ma, QuanGong and Hao, ChaoLong and Yang, XuKui and Qian, LongLong and Zhang, Hao and Si, NianWen and Xu, MinChen and Qu, Dan},
   Title = {Continuous evolution for efficient quantum architecture search},
   Journal = {EPJ Quantum Technology},
   Volume = {11},
   pages = {541},
   DOI = {10.1140/epjqt/s40507-024-00265-7},
   Year = {2024} }

@article{
REF12,
   Author = {Li, Shaochun and Cui, Junzhi and Ren, Jingli},
   Title = {Hybrid classical-quantum neural networks enhanced by quantum architecture search for coronary artery stenosis detection},
   Journal = {Neurocomputing},
   Volume = {618},
   pages = {129111},
   DOI = {10.1016/j.neucom.2024.129111},
   Year = {2024} }

@article{
REF13,
   Author = {Holmes, Zoe and Sharma, Kunal and Cerezo, M. and Coles, Patrick J.},
   Title = {Connecting Ansatz Expressibility to Gradient Magnitudes and Barren
   Plateaus},
   Journal = {PRX Quantum},
   Volume = {3},
   pages = {010313},
   DOI = {10.1103/PRXQuantum.3.010313},
   Year = {2022} }

@article{
M2,
   Author = {Leal, Daivid V. and Araujo, Israel F. and Da Silva, Adenilton J.},
   Title = {Training and meta-training an ensemble of binary neural networks with quantum computing},
   Journal = {Neurocomputing},
   Volume = {572},
   pages = {127169},
   DOI = {10.1016/j.neucom.2023.127169},
   Year = {2024} }

@article{
40,
   Author = {Jerbi, Sofiene and Fiderer, Lukas J. and Poulsen Nautrup, Hendrik and
   Kuebler, Jonas M. and Briegel, Hans J. and Dunjko, Vedran},
   Title = {Quantum machine learning beyond kernel methods},
   Journal = {Nat. Commun.},
   Volume = {14},
   issue = {1},
   Pages = {517},
   DOI = {10.1038/s41467-023-36159-y},
   Year = {2023} }

@inproceedings{
3,
   Author = {Chen, Samuel Yen Chi},
   Title = {Differentiable Quantum Architecture Search in Asynchronous Quantum Reinforcement Learning},
   BookTitle = {2024 IEEE International Conference on Quantum Computing and Engineering, QCE},
   Address= {Montreal, Canada},
   Pages = {1516-1524},
   Year = {2024} }

@article{
REF14,
   Author = {He, Zhimin and Wei, Jiachun and Chen, Chuangtao and Huang, Zhiming and Situ, Haozhen and Li, Lvzhou},
   Title = {Gradient-based optimization for quantum architecture search},
   Journal = {Neural Networks},
   Volume = {179},
   pages = {106508},
   DOI = {10.1016/j.neunet.2024.106508},
   Year = {2024} }

@inproceedings{
D2,
   Author = {Sun, Yize and Liu, Jiarui and Ma, Yunpu and Tresp, Volker},
   Title = {Differentibble Quantum Architecture Search for Job Shop Scheduling Problem},
   BookTitle = {2024 IEEE International Conference on Acoustics, Speech and Signal Processing, ICASSP 2024},
   Address= {Seoul, South Korea},
   Pages = {236-240},
   Year = {2024} }

@article{
RNN,
   Author = {Wang, Gang and Wang, Bang Hai and Fei, Shao Ming},
   Title = {An RNN-policy gradient approach for quantum architecture search},
   Journal = {	Quantum Inf. Process.},
   Volume = {23},
   pages = {1845},
   DOI = {10.1007/s11128-024-04393-y},
   Year = {2024} }

@article{
REF15,
   Author = {Situ, Haozhen and He, Zhimin and Zheng, Shenggen and Li, Lvzhou},
   Title = {Distributed quantum architecture search},
   Journal = {Phys. Rev. A},
   Volume = {110},
   pages = {022403},
   DOI = {10.1103/PhysRevA.110.022403},
   Year = {2024} }

@inproceedings{
REF,
   Author = {Martyniuk, Darya and Jung, Johannes and Paschke, Adrian},
   Title = {Quantum Architecture Search: A Survey},
   BookTitle = {2024 IEEE International Conference on Quantum Computing and Engineering, QCE},
   Address= {Montreal, Canada},
   Pages = {1695-1706},
   Year = {2024} }

@article{
REF16,
   Author = {Soloviev, Vicente P. and Dunjko, Vedran and Bielza, Concha and Larranaga, Pedro and Wang, Hao},
   Title = {Trainability maximization using estimation of distribution algorithms assisted by surrogate modelling for quantum architecture search},
   Journal = {EPJ Quantum Technology},
   Volume = {11},
   pages = {691},
   DOI = {10.1140/epjqt/s40507-024-00282-6},
   Year = {2024} }

@inproceedings{
RL4,
   Author = {Ostaszewski, Mateusz and Trenkwalder, Lea M. and Masarczyk, Wojciech and
   Scerri, Eleanor and Dunjko, Vedran},
   Title = {Reinforcement learning for optimization of variational quantum circuit
   architectures},
   BookTitle = {35th Annual Conference on Neural Information Processing Systems, NeurIPS 2021},
   Address= {Electr Network},
   volume = {34},
   Year = {2021} }

@article{
NA2,
   Author = {Lourens, Matt and Sinayskiy, Ilya and Park, Daniel K. and Blank, Carsten and Petruccione, Francesco},
   Title = {Hierarchical quantum circuit representations for neural architecture search},
   Journal = {NPJ Quantum Information},
   Volume = {9},
   pages = {791},
   DOI = {10.1038/s41534-023-00747-z},
   Year = {2023} }

@inproceedings{
D3,
   Author = {Wu, Wenjie and Yan, Ge and Lu, Xudong and Pan, Kaisen and Yan, Junchi},
   Title = {QuantumDARTS: Differentiable Quantum Architecture Search for Variational Quantum Algorithms},
   BookTitle = {International Conference on Machine Learing, ICML 2023},
   Address = {Honolulu, HI},
   volume = {202},
   Year = {2023} }

@article{
GNN,
   Author = {He, Zhimin and Zhang, Xuefen and Chen, Chuangtao and Huang, Zhiming and Zhou, Yan and Situ, Haozhen},
   Title = {A GNN-based predictor for quantum architecture search},
   Journal = {Quantum Inf. Process.},
   Volume = {22},
   pages = {1282},
   DOI = {10.1007/s11128-023-03881-x},
   Year = {2023} }

@article{
EA1,
   Author = {Li, Yangyang and Liu, Ruijiao and Hao, Xiaobin and Shang, Ronghua and Zhao, Peixiang and Jiao, Licheng},
   Title = {EQNAS: Evolutionary Quantum Neural Architecture Search for Image Classification},
   Journal = {Neural Networks},
   Volume = {168},
   Pages = {471-483},
   DOI = {10.1016/j.neunet.2023.09.040},
   Year = {2023} }

@article{
33,
   Author = {Larocca, Martin and Thanasilp, Supanut and Wang, Samson and Sharma,
   Kunal and Biamonte, Jacob and Coles, Patrick J. and Cincio, Lukasz and
   Mcclean, Jarrod R. and Holmes, Zoe and Cerezo, M.},
   Title = {Barren plateaus in variational quantum computing},
   Journal = {	Nat. Rev. Phys},
   Volume = {7},
   Pages = {174-189},
   DOI = {10.1038/s42254-025-00813-9},
   Year = {2025} }

@article{
EA3,
   Author = {Huang, Yuhan and Li, Qingyu and Hou, Xiaokai and Wu, Rebing and Yung,
   Man-Hong and Bayat, Abolfazl and Wang, Xiaoting},
   Title = {Robust resource-efficient quantum variational ansatz through an
   evolutionary algorithm},
   Journal = {Phys. Rev. A},
   Volume = {105},
   pages = {052414},
   DOI = {10.3390/e25010093},
   Year = {2022} }

@article{
D5,
   Author = {Zhang, Shi Xin and Hsieh, Chang Yu and Zhang, Shengyu and Yao, Hong},
   Title = {Differentiable quantum architecture search},
   Journal = {Quantum Sci. Technol.},
   Volume = {7},
   pages = {0450234},
   DOI = {10.1088/2058-9565/ac87cd},
   Year = {2022} }

@inproceedings{
EA4,
   Author = {Ding, Li and Spector, Lee},
   Title = {Evolutionary Quantum Architecture Search for Parametrized Quantum Circuits},
   BookTitle = {Proceedings of The 2022 Genetic and Evolutionary Computation Conference Companion, GECCO 2022},
   Address= {Boston, MA},
   Pages = {2190-2195},
   Year = {2022} }

@article{
EA5,
   Author = {Du, Yuxuan and Huang, Tao and You, Shan and Hsieh, Min-Hsiu and Tao, Dacheng},
   Title = {Quantum circuit architecture search for variational quantum algorithms},
   Journal = {NPJ Quantum Information},
   Volume = {8},
   pages = {621},
   DOI = {10.1038/s41534-022-00570-y},
   Year = {2022} }

@article{
M4,
   Author = {He, Zhimin and Chen, Chuangtao and Li, Lvzhou and Zheng, Shenggen and Situ, Haozhen},
   Title = {Quantum Architecture Search with Meta-Learning},
   Journal = {Adv. Quantum Technol.},
   Volume = {5},
   pages = {21001348},
   DOI = {10.1002/qute.202100134},
   Year = {2022} }

@article{
EA6,
   Author = {da Silva, Adenilton Jose and Ludermir, Teresa Bernarda and de Oliveira,
   Wilson Rosa},
   Title = {Quantum perceptron over a field and neural network architecture
   selection in a quantum computer},
   Journal = {Neural Networks},
   Volume = {76},
   Pages = {55-64},
   DOI = {10.1016/j.neunet.2016.01.002},
   Year = {2016} }

@article{ 
ZZC,
Author = {Meng, FanXu and Li, ZeTong and Yu XuTao and Zhang, ZaiChen},
Title = {Quantum algorithm for MUSIC-based DOA estimation in hybrid MIMO systems},
Journal = {Quantum Sci. Technol.},
Year = {2022},
Volume = {7},
Pages = {025002},
DOI = {10.1088/2058-9565/ac44dd},
}

@article{ 
E-QAS,
Author = {Linghu, Kehuan and Qian, Yang and Wang, Ruixia and Hu, Meng-Jun and Li,
   Zhiyuan and Li, Xuegang and Xu, Huikai and Zhang, Jingning and Ma, Teng
   and Zhao, Peng and Liu, Dong E. and Hsieh, Min Hsiu and Wu, Xingyao and
   Du, Yuxuan and Tao, Dacheng and Jin, Yirong and Yu, Haifeng},
Title = {Quantum Circuit Architecture Search on a Superconducting Processor},
Journal = {Entropy},
Year = {2024},
Volume = {26},
Pages = {1025},
DOI = {10.3390/e26121025},
}

@inproceedings{ quantumNAS,
Author = {Wang, Hanrui and Ding, Yongshan and Gu, Jiaqi and Lin, Yujun and Pan,
   David Z. and Chong, Frederic T. and Han, Song},
Title = {QuantumNAS: Noise-Adaptive Search for Robust Quantum Circuits},
Booktitle = {2022 IEEE International Symposium on High Performance Computer Architecture, HPCA},
Year = {2022},
Pages = {692-708},
Address= {Electr Network},
DOI = {10.1109/HPCA53966.2022.00057},
}

\end{document}